\documentclass[%
 reprint,
 amsmath,amssymb,
 aps,
]{revtex4-1}
\usepackage{latexsym,amsmath,amssymb,amsfonts,cancel}
\usepackage{graphicx}
\usepackage[utf8]{inputenc}
\usepackage{dcolumn}
\usepackage{bm}

\usepackage[pdftex]{hyperref}
\hypersetup{colorlinks,%
citecolor=blue,%
linkcolor=blue}
\begin{document}

\preprint{APS/123-QED}

\title{Entanglement Generalization in Coupled Harmonic Oscillators}

\author{D. P\'erez-Ad\'an}
 \altaffiliation[]{adan@instec.cu}
\author{F. Guzm\'an}%
\author{O. Rodr\'iguez}%
\affiliation{%
Nuclear Physics Department, Higher Institute of Technologies  and Applied Sciences (InSTEC), Havana, Cuba.}%

\date{\today}

\begin{abstract}
A general and in principle exact approach for the continuous variable entanglement in a system of coupled harmonic oscillators in contact with a thermal bath is formulated. This allows a generalization to describe entanglement's existence between two sites in any system of this kind. It is employed a method of simultaneous diagonalization of two quadratic forms to obtain the uncoupled quantum Hamiltonian. Making use of the transformations that uncouple the system, the covariance matrix for two positions is constructed, and through the positive partial transpose criterion (PPT-criterion), the condition that determines the existence or not of quantum correlation between the sites is obtained. 
\end{abstract}

\maketitle


\section{\label{sec:one}Introduction}
Entanglement is a purely quantum phenomenon that does not have classical analogue and its nature is due to Quantum Mechanics (QM) description of the physical systems. Since the early beginning of the development of the QM theory, the fact that global states describing a compound quantum system can not be expressed as individual products of the subsystem states, put forward the doubt that QM cannot describe elements of the physical reality; all this under the assumption than “realism”, “locality” and “free will” take place in our universe. With these convictions John Stewart Bell \cite{Bell1987} constructed a mathematical apparatus that would prove the existence of hidden variables in QM obtaining some inequalities, the so called “Bell inequalities”, which should not have been violated. However, for surprise of many, were found very well known states that violate these inequalities and this offer to QM, one more time, the confidence vote. \\
The importance of the entanglement theory for quantum information science is well known. On the other hand, it is of interest not only in this field of research. Many physical models are treated as systems of harmonic oscillators, for which revealing the entanglement properties of systems like these has been the inspiration leading to many publications. The mathematical elegance of the harmonic oscillator solution in the different theories and physical formulations has been useful in describing a variety of physical situations occurring in the real world, including those that goes beyond of the mechanical models, for instance, the most recent field theories.\\
The oscillators systems inside the theory of quantum correlations are associated to continuous variables entanglement, which is placed on basis of infinite dimensional Hilbert spaces. Several works \cite{Audenaert2002, Anders2008} have treated the entanglement properties of chains conformed by oscillators obeying different statistical distributions, for instance, the canonical ensemble. In almost all these discussions is common to use identical oscillators with periodic boundary conditions, which have strong symmetry properties. This demand to a real systems is very restrictive and could not be the best choice.\\ That is the reason why this work is focused on performing a generalization of two-site entanglement properties for any system of coupled harmonic oscillators in canonical ensemble. With this approach one is able to predict the existence of quantum correlations between two positions for any configuration in a system of coupled oscillators.\\
This paper is organized as follow. In Section II, the characteristics of a system of coupled quantum oscillators in contact with a thermal bath is treated. Firstly, we will recall the form of the density operator (canonical ensemble) for the harmonic oscillator in different representations and the utility of these to obtain particular mean values, which will be used for the construction of the covariance matrix. Next, we will develop a procedure to uncouple any quantum Hamiltonian of coupled oscillators system, being able to apply the results of the single-particle oscillator to each uncoupled mode. In Section III, with help of the transformation relations obtained in the simultaneous diagonalization process, we will obtain the elements of the covariance matrix; and we will apply a well-known separability criterion, of necessary and sufficient condition, for this kind of states (Gaussian states). Furthermore, we will allude to some particular cases that already have been studied and we will verify the concordance with our generalization. Finally, we will give conclusion and perspective of our work.

\section{\label{sec:two}Coupled Harmonic Oscillators in Canonical Ensemble}

\subsection{\label{sec:twoA}Density operator in canonical ensemble}

The description of continuous variable entanglement for Gaussian states is achieved with help of the covariance matrix \cite{Ryszard2009,Guhne2007,Oleg2009}. To build this matrix for two sites, we should obtain correlations between some magnitudes of the oscillators in those positions of the chain. For this, is very useful to uncouple the oscillators system and obtaining by means of the transformation relations any magnitude of the real configuration. The advantage of uncoupling the system is that each normal mode can be treated as an independent oscillator.\\  
Now we recall some mean values of the one-dimensional quantum harmonic oscillator in canonical ensemble. The single-particle Hamiltonian is
\begin{equation}
\hat{H}(q,p)=\frac{\hat{p}^2}{2m}+\frac{mw^2\hat{q}^2}{2}.  
\label{eq:1}
\end{equation}  
For a canonical ensemble the density operator $\hat{\rho}$ in energy representation can be written \cite{Walter1995}
\begin{gather}
\langle n\mid\widehat{\rho}\mid n'\rangle = \rho_n\delta_{nn'}\hspace*{1cm} n=0,1,2...\nonumber \\
\label{2}\\
\rho_n=[2Senh(\frac{\beta\hbar w}{2})]\cdotp Exp[-\beta\hbar w(n+{1\over 2})].\nonumber
\end{gather}
From similar way to the exposed in \cite{Walter1995} or using the Feynman's path integral method \cite{Feynman1965}, it is possible to obtain the density operator in coordinate or momentum representation. Taking only the diagonal elements ($q’ = q$) one obtains
\begin{equation}
\fontsize{9.5}{0}
\rho(q)= [{mw\over \pi\hbar}Tanh(\frac{\beta\hbar w}{2})]^{{1\over 2}}Exp[-{mw\over \hbar}Tanh(\frac{\beta\hbar w}{2})q^2].
\label{eq:3}
\end{equation} 
Just as well, if we want to obtain the operator in momentum representation, we only would have to change in the previous equation $mw \rightarrow 1/mw$, due to the symmetry of the Hamiltonian to pass from a representation to another one. These different forms are important at the time of calculating mean values, for example, if we want to get the mean value of a function of $\hat{q}$ we use the coordinate representation and analogously the momentum representation for any function of $\hat{p}$. Making some simple calculation we obtain the following mean values    
\begin{gather}
\langle \hat{q}\rangle =0\hspace*{1cm}\langle \hat{q}^2 \rangle = {\hbar\over 2mw}Coth(\frac{\beta\hbar w}{2})\nonumber \\
\label{4}\\
\langle \hat{p}\rangle =0
\hspace*{1cm}\langle \hat{p}^2 \rangle = {\hbar wm\over 2}Coth(\frac{\beta\hbar w}{2}).\nonumber
\end{gather}
These calculated magnitudes will be important at the time of constructing the elements of covariance matrix for two positions of the chain, since in the following subsection we will apply a procedure to transform the Hamiltonian of the coupled system, to one of uncoupled oscillators. After having done this, it will be necessary to make use of the values calculated in Eq. \eqref{4} for individual oscillators.

\subsection{\label{sec:twoB}System of coupled harmonic oscillators}
The diagonalization of the Hamiltonian of a coupled oscillators system has a great importance as much in classical mechanics as in QM, since this process passes to a representation of completely decoupled oscillators. But this method has a significant difference in relation to both theories; in classical mechanics the relation between momentum and position is linear, however, in QM these are operators that obey the Heisenberg relation. This fact has some consequences in the diagonalization process of the Hamiltonian. Any system's Hamiltonian (classic or quantum) of this kind can be represented as a sum of two quadratic forms, one in the momentum variables and another one in position variables. Passing to so-call normal modes means to do a simultaneous diagonalization of those quadratic forms, taking into account the interrelation between the variables. Now we will propose a procedure to carry out this diagonalization, equivalent to a Bogoliubov transformation, but in position-momentum representation.
Let us consider a system of one-dimensional coupled oscillators; the quantum Hamiltonian of the system ($H_S$) is
\begin{equation}
H_S (\vec{\hat{q}},\vec{\hat{p}})= {1\over 2}\vec{\hat{p}}^T\hat{\textbf T}\vec{\hat{p}}+{1\over 2}\vec{\hat{q}}^T\hat{\textbf V}\vec{\hat{q}},
\label{eq:5}
\end{equation}
where $ \vec{\hat{p}} $ and $ \vec{\hat{q}} $ are single-column matrices whose elements are the momentum operators ($\hat{p}_i$) and the position operators ($\hat{q}_i$) of each oscillator respectively, $ \hat{\textbf T} $ is a $n\times n$ diagonal matrix associated to the kinetic energy part of the system and $\hat{\textbf V}$ is a $n\times n$ symmetric matrix associated to the potential energy part of the system. The elements of these matrices correspond to the characteristics of the oscillators and to the way that couple. The superscript $T$ is for transpose, $*$ for conjugate, and $\dagger$ for Hermitian adjoint (conjugate transpose). In order to be able to uncouple the oscillators, is necessary to find a set of new variables ($\hat{Q}_i,\hat{P}_i ; i=1…n$) lineally related with the originals ($\hat{q}_i,\hat{p}_i ; i=1…n$), in such a way that this transformation diagonalizes simultaneously the matrices $ \hat{\textbf T} $ and $ \hat{\textbf V} $; but this is not enough, the new variables (operators) have to continue fulfilling the Heisenberg commutation relation \cite{Sakurai1994}. That's to say
\begin{equation}
[\hat{q}_i,\hat{p}_j]=[\hat{Q}_i,\hat{P}_j]=i\hbar\hat{\rm 1}\delta_{ij},
\label{eq:6}
\end{equation}
where $[{\hspace{0.2cm}} , {\hspace{0.2cm}}]$ represent the commutator and $ \hat{\rm 1} $ the identity operator. We denote by $ \hat{\textbf A} $ the matrix that accomplishes the transformation from $\hat{q}_i$ to $\hat{Q}_i$; then, it can be proven with help of Eq. \eqref{eq:6} that the transformations have to hold
\begin{equation}
\vec{\hat{q}}=\hat{\textbf A}\vec{\hat{Q}} {\hspace{0.8cm}}\Longrightarrow {\hspace{0.8cm}} \vec{\hat{p}}=(\hat{\textbf A}^{-1})^T\vec{\hat{P}}.
\label{eq:7}
\end{equation}  
Now the task is finding that matrix, in such a way that when the transformations are applied, the Hamiltonian becomes diagonal. For this, first let us define the following diagonal and dimensionless matrix, denoted by $ \hat{\textbf R} $
\begin{equation}
\hat{\textbf R}=\sqrt{{\hat{\textbf T}}\over \mu} {\hspace{2cm}} \large{ \mu =\sqrt[n]{\small{\textbf {Det} [\hat{\textbf T}]}}},
\fontsize{10}{0}
\label{eq:8}
\end{equation}
and let us insert the following intermediate change of variable
\begin{equation}
\vec{\hat{q}}=\hat{\textbf R}\vec{\hat{q'}}{\hspace{2.2cm}}\vec{\hat{p}}=\hat{\textbf R}^{-1}\vec{\hat{p'}}.
\label{eq:9}
\end{equation}
This transformation fulfills Eq. \eqref{eq:7}, so that the new operators $ \hat{q_i'} $ and $ \hat{p_i'} $ satisfy the Heisenberg commutation relation. Applying this transformation to the Hamiltonian Eq. \eqref{eq:5} we obtain
\begin{equation}
H_S (\vec{\hat{q'}},\vec{\hat{p'}})= {1\over 2}\mu \vec{\hat{p'}}{}^{T}\hat{\textbf I}\vec{\hat{p'}}+{1\over 2}\vec{\hat{q'}}{}^{T}\hat{\textbf R}\hat{\textbf V}\hat{\textbf R}\vec{\hat{q'}},
\label{eq:10}
\end{equation} 
being $ \hat{\textbf I} $ the $n\times n$ identity matrix. Performing the next transformation (this also fulfills Eq. \eqref{eq:7})
\begin{equation}
\vec{\hat{q'}}=\hat{\textbf S}\vec{\hat{Q}}{\hspace{2.2cm}}\vec{\hat{p'}}=\hat{\textbf S}\vec{\hat{P}},
\label{eq:11}
\end{equation} 
where $ \hat{\textbf S} $ is the orthogonal matrix of eigenvector of $ \hat{\textbf R}\hat{\textbf V}\hat{\textbf R} $, mathematically
\begin{equation}
\hat{\textbf S}^T\hat{\textbf R}\hat{\textbf V}\hat{\textbf R}\hat{\textbf S}=\hat{\textbf D},
\label{eq:12}
\end{equation}
where the elements of the diagonal matrix $\hat{\textbf D}(\lambda_i)$ are the roots of the equation
\begin{equation}
\textbf{Det}[\hat{\textbf R}\hat{\textbf V}\hat{\textbf R}-\lambda \hat{\textbf I}]=0,
\label{eq:13}
\end{equation}
and inserting \eqref{eq:11} into \eqref{eq:10}, with help of Eqs. \eqref{eq:12} and \eqref{eq:13} the Hamiltonian becomes
\begin{align}
H_S (\vec{\hat{Q}},\vec{\hat{P}})& =  {1\over 2}\mu\vec{\hat{P}}^T\hat{\textbf I}\vec{\hat{P}}+{1\over 2}\vec{\hat{Q}}^T\hat{\textbf D}\vec{\hat{Q}}\nonumber \\ 
& = \sum_{i=1}^{n}[{1\over 2}\mu \hat{P}_i^2+{1\over 2}\lambda _i \hat{Q}_i^2].
\label{eq:14}
\end{align} 
Using Eqs. \eqref{eq:9} and \eqref{eq:11} is easy to see that the matrix $\hat{\textbf A}$ wanted is
\begin{equation}
\hat{\textbf A}=\hat{\textbf R}\hat{\textbf S}.
\label{eq:15}
\end{equation}
In this way has been converted the original Hamiltonian Eq. \eqref{eq:5} to a sum of Hamiltonians of uncoupled oscillators Eq. \eqref{eq:14}, in such a way that each one can be resolute individually. It is easy to notice from Eq. \eqref{eq:14} that the frequencies ($\omega_{i}^f; i=1…n$) of the vibrational modes of the lattice, the phonons, are given by
\begin{equation}
\omega_{i}^f=\sqrt{\mu\lambda_i},
\label{eq:16}
\end{equation}
and it can be proven that these are the same obtained in the classical procedure.

\subsection{\label{sec:twoC}General Density Operator}

For a system of coupled oscillators in a thermal bath, once diagonalized the Hamiltonian, we can write the density operator of the system as the tensor product of the density operators associated to each normal mode; in other words, the general state is separable. This operator expressed in the new set of coordinates is
\begin{equation}
\rho (\vec{{Q}})=\prod_{i=1}^n\rho_i ({Q_i}),
\label{eq:17}
\end{equation}
with
\begin{gather}
\rho_i ({Q_i})=\Big([{\sqrt{\lambda _i}\over \sqrt{\mu}\pi\hbar}Tanh(\frac{\beta\hbar\sqrt{\mu\lambda _i}}{2})]^{\displaystyle {1\over 2}}\Big)\nonumber \\
\hspace{1.5cm}\times\hspace{0.5cm} \Big(Exp[-{\sqrt{\lambda _i}\over \sqrt{\mu}\hbar}Tanh(\frac{\beta\hbar\sqrt{\mu\lambda _i}}{2})Q_i^2]\Big).
\label{18}
\end{gather}
From similar way to the exposed in Section II, it can be obtained the mean values of the magnitudes $\langle \hat{Q}_i \rangle, \langle \hat{Q}_i^2 \rangle $, $\langle \hat{P}_i \rangle$ and $\langle \hat{P}_i^2 \rangle $. In this case, each mean value depends exclusively of the individual characteristics of the oscillator that it represents, because when is calculated the mean value of the $i^{th}$ magnitude, the remainder integrals that do not contain $i^{th}$ term reduce to unity. Then
\begin{gather}
\langle \hat{Q_i}\rangle =0 \hspace{1cm}
\langle \hat{Q_i}^2 \rangle = {\hbar\sqrt{\mu}\over 2\sqrt{\lambda _i}}Coth(\frac{\beta\hbar\sqrt{\mu\lambda_ i}}{2}) \nonumber \\
\label{19}\\
\langle \hat{P_i}\rangle =0{\hspace{1cm}} \langle \hat{P_i}^2 \rangle = {\hbar\sqrt{\lambda_ i}\over 2\sqrt{\mu}}Coth(\frac{\beta\hbar\sqrt{\mu\lambda_ i}}{2}).\nonumber
\end{gather}      
Now already is possible to construct the covariance matrix for two sites and applying the separability criterion.

\section{\label{sec:three}Two-Site Entanglement}

\subsection{\label{sec:threeA}Covariance matrix}

As we want to obtain the two-site entanglement condition of a harmonic lattice, is convenient to build the covariance matrix; this is a real, symmetric and positive matrix which reveals the occurrence of entanglement between two oscillators of the chain. This matrix is defined in \cite{Ryszard2009,Anders2008} and is given by 
\begin{equation}
\hat{\Gamma}_{ij}(\rho)=\langle \hat{R}_i\hat{R}_j+\hat{R}_j\hat{R}_i \rangle -2\langle \hat{R}_i\rangle\langle \hat{R}_j \rangle ,
\label{eq:20}
\end{equation}
where $\hat{R}_i$ are the components of the vector: $\vec{\hat{R}}$ =($ \hat{q}_1, \hat{p}_1, \hat{q}_2, \hat{p}_2,…,  \hat{q}_n, \hat{p}_n $), in which $\hat{q}_i$ and $ \hat{p}_i$ denote the position and momentum operators respectively of the $i^{th}$ oscillator. For two sites in the chain, this is a $4\times 4$ matrix whose elements will be denoted by letters of the alphabet. In order to construct this, we will work with Eqs. \eqref{eq:7} and \eqref{19}; furthermore, we will apply the linear properties of the mean values. Using those equations it is easy to notice that $ \langle\hat{q}_i\rangle =\langle\hat{p}_i\rangle =0  $. The terms $\langle\hat{q}_i\hat{p}_i+\hat{p}_i\hat{q}_i\rangle$ cancel out for the following reason: for the harmonic oscillator in canonical ensemble the mean value of `position multiplied by momentum' (i.e. $ \widehat{qp}=\widehat{pq}={\hat{q}\hat{p}+\hat{p}\hat{q}\over 2} $) is zero. This can be proven demonstrating that the average $\langle \widehat{qp}\rangle$ takes the value zero for all the eigenstates of the Hamiltonian Eq. \eqref{eq:1}, and as a result, also for a canonical distribution of one-dimensional harmonic oscillators. Afterwards, this average cancels out for each uncoupled mode, and as a consequence, for the operators ($\hat{q}_i,\hat{p}_i$) too. Now operating with Eqs. \eqref{eq:7} and \eqref{19} we obtain the following expressions
\begin{gather}
\langle\hat{q}_i\hat{q}_j+\hat{q}_j\hat{q}_i\rangle =\hbar\sqrt{\mu}\sum_{l=1}^n A_{il}A_{jl}{Coth(\frac{1}{2}\beta\hbar\sqrt{\mu\lambda_ l)}\over \sqrt{\lambda_ l}}\nonumber \\
\label{21}\\
\langle\hat{p}_i\hat{p}_j+\hat{p}_j\hat{p}_i\rangle ={\hbar\over \sqrt{\mu}}\sum_{l=1}^n A_{li}^{-1}A_{lj}^{-1}\sqrt{\lambda_ l}Coth(\frac{\beta\hbar\sqrt{\mu\lambda_ l}}{2}),\nonumber
\end{gather}     
where $A_{il}$ represent the element ($i,l$) of the matrix $ \hat{\textbf A} $ and $A_{il}^{-1}$ the element ($i,l$) of the inverse of this matrix. In case of being $ \hat{\textbf A} $ a complex matrix, we have to take the real part in the previous expressions. With this, the covariance matrix takes the form
\begin{equation}
\hat{\pmb{\Gamma}}_{ij}(\rho) = \left( \begin{array}{cccc}
A & 0 & C & 0\\
0 & E & 0 & G\\
C & 0 & H & 0 \\
0 & G & 0 & J
\end{array}
\right),
\label{eq:22}
\end{equation}
with
\begin{eqnarray}
A=2\langle \hat{q}_i^2 \rangle {\hspace{0.8cm}} E=2\langle \hat{p}_i^2 \rangle {\hspace{0.8cm}}
C=\langle\hat{q}_i\hat{q}_j+\hat{q}_j\hat{q}_i\rangle \nonumber\\
\\
H=2\langle \hat{q}_j^2 \rangle {\hspace{0.8cm}} J=2\langle \hat{p}_j^2 \rangle {\hspace{0.8cm}}
G=\langle\hat{p}_i\hat{p}_j+\hat{p}_j\hat{p}_i\rangle , \nonumber
\label{eq:23}
\end{eqnarray}
being $A$, $E$, $H$, $J$ particular cases from Eq. \eqref{21}.

\subsection{\label{sec:threeB}Separability criterion. Logarithmic negativity}

As it was shown in \cite{Anders2008}, for Gaussian continuous variable states of two modes there exist a necessary and sufficient separability criterion. It is a variation of the positive partial transpose criterion (PPT-criterion) for discrete systems \cite{Pawel1997,Ryszard2009}. For separability of two sites i and j in the chain the criterion requires the positivity of the matrix inequality \cite{Anders2008},
\begin{equation}
\hat{\rho}_{ij} {\hspace{0.2cm}} separable {\hspace{0.7cm}} \Longleftrightarrow {\hspace{0.7cm}} \hat{\pmb{\Gamma}}_{ij}^{Tp}(\rho)+i\hbar\bigoplus_{i,j}\hat{\pmb{\sigma}}\geq 0, 
\label{eq:24}
\end{equation}   
where {\hspace{0.1cm}}$ \hat{\pmb{\sigma}} = \left( \begin{array}{cc}
0 & 1\\
-1 & 0
\end{array}
\right)
 $ and $ \hat{\pmb{\Gamma}}_{ij}^{Tp}(\rho) $ is given by ($ \hat{\textbf{I}}_{2} \equiv \hat{\textbf{I}}_{2\times 2} $)
 \begin{equation}
\hat{\pmb{\Gamma}}_{ij}^{Tp}(\rho)=\Big[\hat{\textbf{I}}_{2}\oplus\left(\begin{array}{cc}
1 & 0\\
0 & -1
\end{array}
\right)\Big]\hat{\pmb{\Gamma}}_{ij}\Big[\hat{\textbf{I}}_{2}\oplus\left(\begin{array}{cc}
1 & 0\\
0 & -1
\end{array}
\right)\Big]. 
\label{eq:25}
\end{equation}
It is easy to check that in this case, to obtain $ \hat{\pmb{\Gamma}}_{ij}^{Tp}(\rho) $ we simply replace $G$ by $–G$. Now to decide the positiveness of the matrix \eqref{eq:24} we will apply the necessary and sufficient condition of positivity of all its principal minors \cite{Strang1988,Kurosh1968}. The first determinant (upper left $1 \times 1$ matrix) is obviously positive since $A$ is a positive defined magnitude. The second determinant (upper left $2 \times 2$ matrix) is positive too, and it can be proven with help of the expressions in Eq. \eqref{21} for $A$ and $E$, and then applying Cauchy-Schwarz-Buniakowsky inequality \cite{Alan2008}. Also with help of this inequality it is possible to demonstrate the positiveness of other expressions, as the exposed in \cite{Isar2010}. Then, making use of the obtained inequalities in \cite{Isar2010}, it can be tested that the positiveness of the third determinant (upper left $3 \times 3$ matrix) is totally related to the fourth principal minor's positiveness. Specifying better; it is demonstrated that if the third minor is negative, the fourth minor (whole matrix) is also negative, and the opposite, if the fourth minor is positive, it will be the third one too. Then is enough with knowing the behaviour of the determinant of \eqref{eq:24} and that way we will know if this matrix is positive or not. With this the entanglement criterion can be written of the following way, where has been taken into account $\hbar =1$:
\begin{equation}
L=(C^2-AH)(G^2-EJ)+2CG-HJ-AE+1 \geq 0.
\label{eq:26}
\end{equation}          
If that inequality is violated, then the sites “$i$” and “$j$” of the chain are entangled, and the magnitude of their entanglement can be given by the logarithmic negativity $E_N$, defined in \cite{Ryszard2009,Vidal2002} 
\begin{gather}
\hspace{-3cm}E_N =-{1\over 2}Log_2 \Big \{ {1\over 2}(AE+HJ)-CG\nonumber \\
\hspace{0.5cm}
-\sqrt{[{1\over 2}(AE+HJ)-CG]^2-(C^2-AH)(G^2-EJ)} \Big \}.
\label{27}
\end{gather}
With this treatment is possible to detect the existence of entanglement between two positions of all system of coupled oscillators embedded in a thermal bath. We have treated only the one-dimensional problem here, because we can decompose the multidimensional case in so many one-dimensional systems as dimensions be trying and to solve each one separately.

\subsection{\label{sec:threeC}Two special cases: Linear and circular chain}

Let us analyze two cases of special interest in the construction of approximate models, for structures with certain symmetry. The importance of these two configurations lies in the fact that 
is not required a numerical treatment in the diagonalization process, because there are analytical expressions for the matrix $ \hat{\textbf A} $. \\ Let us suppose a one-dimensional chain of coupled identical harmonic oscillators, of mass $ m $ and frequency $ w $, with periodic boundary conditions (circular chain). As much in this case as in the one of linear chain, the matrix $ \hat{\textbf R} $ is the identity matrix, hence, the matrix that diagonalizes both quadratic forms is the one that diagonalizes $ \hat{\textbf V} $. In this case the matrix associated to the potential energy ($ \hat{\textbf V}_C $) presents the following form (ignoring the term ${1\over 2}mw^2 $ multiplying the matrix)
\begin{equation}
\hat{\textbf{V}}_C = \left( \begin{array}{rrrrrrrr}
2 & -1 & 0 & . & . & . & 0 & -1\\
-1 & 2 & -1 &  &  &  &  & 0\\
0 & -1 & 2 & . &  &  &  & .\\
. &  & . & . & . &  &  & .\\
. &  &  & . & . & . &  & .\\
. &  &  &  & . & . & -1 & 0\\
0 &  &  &  &  & -1 & 2 & -1\\
-1 & 0 & . & . & . & 0 & -1 & 2
\end{array}
\right).
\label{eq:28}
\end{equation}
The matrix $ \hat{\textbf A}_C $ which diagonalizes $ \hat{\textbf V}_C $ (see for instance \cite{Audenaert2002}) is
\vspace{-0.2cm}
\begin{equation}
(A_C)_{kl}={1\over \sqrt{n}}Exp[i{2\pi\over n}kl],
\label{eq:29}
\end{equation}
\vspace{-0.1cm}
converting it into the diagonal matrix
\begin{equation}
(D_C)_{kk}=\lambda_ k = 4Sen^2 ({\pi k\over n}).
\label{eq:30}
\end{equation}
\vspace{-0.1cm}
It is not difficult to check that in this case the covariance matrix becomes independent of the position pair ($i, j$), depending only of the difference between that positions: $i - j$; result expected given the circular symmetry. Furthermore, since $A$ becomes equal to $H$ and $E$ equal to $J$, the inequality \eqref{eq:26} can be factorized in two terms, and therefore, it can be separated in two inequalities. In \cite{Anders2008} a more detailed description of this case (circular chain) can be obtained.\\  
In order to analyze the linear chain structure, we suppose the same previous model, but this time with fixed boundary conditions. Now the matrix $ \hat{\textbf V}_L $ takes a similar aspect to $ \hat{\textbf V}_C $, but instead of “ $-1$ ” in the upper right corner and in the lower left corner (Eq. \eqref{eq:28}), we put there “ $0$ ”. Now the matrix which diagonalizes $ \hat{\textbf V}_L $ is
\begin{equation}
(A_L)_{kl}=\sqrt{{2\over n+1}}Sen[{\pi kl\over n+1}],
\label{eq:31}
\end{equation} 
converting it into the diagonal matrix
\begin{equation}
(D_L)_{kk}=\lambda_ k = 4Sen^2 [{\pi k \over 2(n+1)}].
\label{eq:32}
\end{equation}
In this case the covariance matrix is different for each position pair ($i, j$). It would be interesting to do a comparison between these two configurations, taking into account the negativity for two equally separated positions in the chain and the critical temperature below which entanglement survives.

\section{\label{sec:four}Conclusion}

In this work it has been developed the physical-mathematical basement for entanglement detection in systems of coupled harmonic oscillators obeying a canonical distribution. It was achieved a diagonalization of the Hamiltonian of any quantum oscillators system, considering that this can be represented algebraically as the sum of two quadratic forms, whose variables fulfill the Heisenberg commutation relation. Later it was applied the Peres-Simon necessary and sufficient condition for separability of two-mode Gaussian states, in order to find the relation that determines the existence or not of entanglement. It was confirmed the concordance of this generalization with the results obtained previously in \cite{Anders2008} for the circular uniform chain. In general, it can be argued, that in structures of oscillators the entanglement between two positions is much related with the geometrical configuration of the oscillators system, as well as the interaction strength between these two positions. It would be interesting to accomplish a detail study of how the entanglement depends on the coherence between the components of the chain; it is possible that the loss of coherence (big differences in masses or in frequencies) leads to a weaker correlation.

\end{document}